\documentclass[runningheads]{llncs}

\usepackage{amsfonts}
\usepackage{booktabs}
\usepackage{mathtools}
\usepackage{supertabular}
\usepackage{multirow}
\usepackage{listings}
\usepackage{appendix}
\usepackage{fancyvrb}
\usepackage{fvextra}

\usepackage{graphicx}
\usepackage{amsmath}
\begin{document}
\title{Multimodal Container Planning: a QUBO Formulation and Implementation on a Quantum Annealer}
\titlerunning{Multimodal Planning on a Quantum Annealer}

\author{F. Phillipson\orcidID{0000-0003-4580-7521} \and I. Chiscop}
\institute{TNO, PO Box 96800, 2509 JE The Hague, The Netherlands}

\authorrunning{F. Phillipson and I. Chiscop}

\maketitle              
\begin{abstract}
Quantum computing is developing fast. Real world applications are within reach in the coming years. One of the most promising areas is combinatorial optimisation, where the Quadratic Unconstrained Binary Optimisation (QUBO) problem formulation is used to get good approximate solutions. Both the universal quantum computer as the quantum annealer can handle this kind of problems well. In this paper, we present an application on multimodal container planning. We show how to map this problem to a QUBO problem formulation and how the practical implementation can be done on the quantum annealer produced by D-Wave Systems.

\end{abstract}

\section{Introduction}
In container logistics various transportation concepts are known \cite{steadieseifi2014multimodal}. Unimodal transport usually
refers to loading cargo onto a single transportation mode, usually a truck, and driving it from origin to destination. Multimodal transport refers to transporting cargo from origin to destination by more than one transportation mode. In intermodal transport, again more than one transportation mode may be used, but the containment unit (container) must always be of standardised size. Co-modal transport looks like multimodal transport, but requires a consortium of shippers and has a focus on exploiting the benefits of each transportation mode in a smart way. Finally, synchromodal transport is a version of intermodal transport that focuses on real-time planning flexibility and coordination between different shippers, both using large amounts of real-time data.

When looking at the planning of container flows, it is often categorised on three levels of problems: on strategic, tactical and operational level \cite{van2015synchromodal}. Strategic problems concern long-term investments in the transportation network, for example, where to build new terminals. Tactical problems may concern service design, for example, determining how many times in a month a barge should make a round-trip. Operational problems concern using a current network in an optimal way for problems occurring in the present. Little attention has been on operational problems \cite{mes2016synchromodal}. Operational problems can be divided into various classes: (a) problems in which containers are assigned to existing barge services, (b) problems in which the routes of barges are determined for a given demand, and (c) problems in which both the assignment of containers to barges and the route of barges are decided upon \cite{zweers2019optimizing}.

In an operational, dynamic, environment with a lot of uncertainty, or in a synchromodal environment where real time system characteristics are used in the planning, short computation times are crucial, which make it possible to repeat the planning process with a high frequency. To this end, efficient algorithms are proposed \cite{steadieseifi2014multimodal} or techniques are developed to reduce the complexity of the problem \cite{kalicharan2019reduction}. Next to software and algorithm enhancements to speed up the computation, also hardware developments can play a role. A promising direction here is quantum computing. Note that this will not be a solution for the short term, where the current generation quantum computers is far too small to compete with classical computers and their advanced solvers and algorithms. This work is meant to sketch a new direction for medium and long term solutions.

The past decade has seen the rapid development of the two paradigms of quantum computing, quantum annealing and gate-based quantum computing. In 2011 D-Wave Systems announced the release of the world's first commercial quantum annealer\footnote{\url{https://www.dwavesys.com/news/d-wave-systems-sells-its-first-quantum-computing-system-lockheed-martin-corporation}} operating on a 128-qubit architecture, which has since been continually extended up to the 2048-qubit version, available from 2017\footnote{\url{https://www.dwavesys.com/press-releases/d-wave\%C2\%A0announces\%C2\%A0d-wave-2000q-quantum-computer-and-first-system-order}}. D-Wave announced in 2019 a (more than) 5000 qubit system available mid-2020, using their new Pegasus-technology based chip with 15 connections per qubit\footnote{\url{https://www.dwavesys.com/press-releases/d-wave-previews-next-generation-quantum-computing-platform}}. Even more recently, quantum supremacy is claimed to have been achieved by Google's 54-qubit Sycamore gate-based computer \cite{Google19}. These technological advances have led to a renewed interest in finding classical intractable problems suited for quantum computing.

A particular category of problem that are expected to fit well on quantum devices are (combinatorial) optimisation problems. An important trail in implementing optimisation problems on quantum devices is the Quadratic Unconstrained Binary Optimisation (QUBO) problem. These QUBO problems can be solved by a Universal (gate-based) quantum computer using a QAOA (Quantum Approximate Optimisation Algorithm) implementation \cite{farhi2014quantum} or by a quantum annealer, such as D-Wave \cite{johnson2011quantum}. For already a large number of combinatorial optimisation problems the QUBO representation is known \cite{glover2019tutorial,lucas2014ising}. Well-known examples that have been implemented on one of D-Wave's quantum processors include maximum clique \cite{max_clique}, capacitated vehicle routing \cite{vehicle_routing}, minimum vertex cover \cite{min_vertex}, set cover with pairs \cite{set_cover_with_pairs}, Multi-Service Location Set Covering Problem \cite{quantum_multi_service_loc}, traffic flow optimisation \cite{traffic_flow} and integer factorisation \cite{integer_factorization}. These studies have shown that although the current generation of D-Wave annealers may not yet have sufficient scale, precision and connectivity to allow faster or higher quality solutions, they have the suitable infrastructure for modelling real-world instances of these problems, effectively decomposing these into smaller sub-problems and solving these on a real Quantum Processing Unit (QPU).The QUBO problems can also be solved by simulated (quantum) annealing on a conventional computer. Then, however, no quantum advantage can be expected.

In this paper we propose two QUBO formulations for a container assignment problem and show the implementation on both simulated annealing as well as on the D-Wave QPU. The size of the current generation of quantum computers limits us to small problems, with a limited number of decision variables. Therefor, we propose first a QUBO-formulation that only uses two potential paths for each container, and expand this to four alternatives per container.

\section{Mathematical framework}
In this work a deterministic container-to-mode problem is studied. To give a formal definition the framework of \cite{de2017framework} is used. The following $\bar R | \bar D, [D2R] | social(1)$-problem is meant: to assign freight containers to transportation modes, such that the containers reach their destinations before a deadline against minimum total cost, given that the transport modes have fixed given schedules and all features of the problem are deterministic \cite{huizing2017general}. These problems are often solved using minimum cost multi-commodity flow problems on time-dependent graphs and space-time networks \cite{andersen2009service,ding2008finding}. Finding a non-negative two-commodity flow on a directed graph, however, is proven to be NP-complete \cite{even1975complexity}. Finding a flow with minimum cost and with at least two commodities must be at least as difficult, however, the LP-relaxation of the studied problems almost always has an integral optimum \cite{huizing2017general}. To study the potential of quantum computing for this problem we propose a QUBO formulation. 

\subsection{Approach}
Given the state of the quantum devices, which are quite small at this moment, we will limit the size of the problem by giving each container a limited number of paths through the transportation network, here either 2 or 4. For this we propose the following approach:
\begin{enumerate}
    \item Select a set (2 or 4) of alternative routes per container; In the case of 2 alternatives we take the best multimodal path and the uni-modal (trucking) path. In the case of 4 alternative routes, we take again the uni-modal path and the 3 maximum dissimilar paths.
    \item Define the QUBO-problem.
    \item Solve the QUBO-problem.
    \item Evaluate the solution and go back to step 1 if necessary to select different sets of alternatives.
\end{enumerate}

\subsection{QUBO Formulation}
We now give the definition of the QUBO-problem \cite{glover2019tutorial} and propose the formulation of the QUBO for the two problems. The QUBO is expressed by the optimisation problem:
\begin{equation} \label{eq:1}
\text{QUBO: min/max } y=x^t Qx,
\end{equation}
where $x \in \{0,1\}^n$, the decision variables and $Q$ is a $n \times n$ coefficient matrix. QUBO problems belong to the class of NP-hard problems. Another formulation of the problem, often used, equals 
\begin{equation} \label{eq:2}
\text{QUBO: min/max } H=x^tq + x^tQx,
\end{equation}
or a combination of multiple of these terms 
\begin{equation} \label{eq:3}
\text{QUBO: min/max } H=A \cdot H_A + B \cdot H_B + \cdots,
\end{equation} 
where $A, B, \dots$ are weights that can used to tune the problem and include constraints into the QUBO. We will use this representation in the next sections, where we will present the QUBO formulation for the 2 and 4 alternative route problems.

\subsubsection{Creating a QUBO}
For already a large number of combinatorial optimisation problems the QUBO representation is known \cite{glover2019tutorial,lucas2014ising}. Many constrained integer programming problems can be transformed easily to a QUBO representation. Assume that we have the problem 
\begin{equation} 
\text{min } y=c^t x, \text{ subject to } Ax=b,
\end{equation}
then we can bring the constraints to the objective value, using a penalty factor $\lambda$ for a quadratic penalty:
\begin{equation} 
\text{min } y=c^t x + \lambda (Ax-b)^t(Ax-b).
\end{equation}
Using $P=Ic$, the matrix with the values of vector $c$ on its diagonal, we get
\begin{equation} 
\text{min } y=x^t Px + \lambda (Ax-b)^t(Ax-b)=x^t Px + x^t Rx + d = x^t Qx,
\end{equation}
where matrix $R$ and the constant $d$ follow from the matrix multiplication and the constant $d$ can be neglected, where it does not influence the optimisation problem.

\subsubsection{2-alternative route QUBO}
We look at a situation where containers have to be shipped in an intermodal or synchromodal network. To limit the number of options, we give every container two possible paths through the network, one using a truck only, the other a multimodal or multitrack path through the network. Each track or modality in this network has a specific capacity. We define a QUBO representation for this problem. We assume that all routes are feasible, with regard to due dates etc.
\begin{table}[]
\centering
\begin{tabular}{ll}\toprule
\multicolumn{2}{l}{\textbf{Indices}}   \\ \hline
$i$           & Containers, $i \in \{1, \ldots, n\}$           \\ 
$j$           & Tracks, $j \in \{1, \ldots, m\}$           \\ \hline

\multicolumn{2}{l}{\textbf{Parameters}} \\ \hline
$c_i^b$     & costs for using multimodal route container $i$  \\
$c_i^t$     & costs for using truck by container $i$ \\
$V_j$       & capacity of track $j$ \\
$r_{i}$     & route for container $i$ \\ 
$r_{ij}$     & 
=$ \begin{cases} 
      1 & \text{if route } i \text{ contains track } j \\
      0 & \text{otherwise}
   \end{cases}
$ \\ \hline
\multicolumn{2}{l}{\textbf{Decision variables}} \\ \hline
$x_{i}$     & 
=$ \begin{cases} 
      1 & \text{if container } i \text{ is transported by truck} \\
      0 & \text{if container } i \text{ is transported by barge route } r_i
   \end{cases}
$ \\ \bottomrule
\end{tabular}
\caption{Parameters and decision variables.}
\label{tab:parameters1}
\end{table}

The binary integer linear program for the problem is formulated as follows:
\begin{align}
\min &\sum_{i=1}^n c_i^b + (c_i^t-c_i^b)x_i, \label{eq:4} \\
\text{such that} \quad &\sum_{i=1}^{n} (1-x_i)r_{ij} \leq V_j \quad \forall j \in \{1,\dots m\}. \label{eq:5}
\end{align}
\noindent

Here for every container $i$ it has to be decided whether it is sent using a truck $(x_i=1)$ or using the multitrack multimodal path $(x_i=0)$. Eq. \ref{eq:4} minimises the costs of the choices, where per container $i$ the choice $x_i=0$ leads to costs $c_i^b$ and  $x_i=1$ to $c_i^b + (c_i^t-c_i^b) = c_i^b -c_i^b + c_i^t = c_i^t.$ Eq. \ref{eq:5} makes sure that the capacity constraints of the modalities or tracks are met.

Based on the model in Eq. (\ref{eq:4}-\ref{eq:5}), the QUBO formulation of the problem can also be derived. The overall solution for this QUBO is given by:
\begin{align}
    \min \quad& A\cdot H_{A} + B \cdot H_{B}, \label{eq:6}\\
    \text{ with }
    H_{A} &= \sum_{i=1}^n c_i^b + (c_i^t-c_i^b)x_i, \label{eq:7 }\\
    H_{B} &= \sum_{j=1}^{m} \Big(\sum_{i=1}^{n} (1-x_i)r_{ij} + \sum_{k=0}^{K}2^{k}y_{jk} - V_j \Big)^{2}, \label{eq:8}
\end{align}

where $A$ and $B$ denote penalty coefficients to be applied such that the constraints will be satisfied and $y_{ik}$ denotes additional slack variables. These $K \cdot m $ binary slack variables are necessary to remodel Eq. \ref{eq:5} into equality constraints, as they are required in a QUBO-formulation. The parameter $K$ follows from $K=\max_j \big( \log_2 (V_j)\big)$. When determining the penalty coefficients, we can set $A=1$ and look for a good value for $B$. Rule of thumb is that the gain of violating a constraint must be lower than the costs. This means in this problem that $B>\max_i c^b_i$.

Note that if capacity is bounding for all tracks, Eq. \ref{eq:5} changes to an equality, which removes the need for the slack variables in Eq. \ref{eq:8}. If the capacity is not bounding for all tracks, i.e.., the capacity is very high, Eq. \ref{eq:5} and Eq. \ref{eq:8} disappear from the problem for certain values of $j$, instead of allowing the slack variables for very high values. 

\subsubsection{4-alternative route QUBO}
If we give the problem, next to trucking, three multitrack and multimodal options to choose from, we get the following model. Now we use an extra binary variable to represent this choice: $\{11\}$ stands for trucking, $\{01\}$ for route 1, $\{10\}$ for route 2 and $\{00\}$ for route 3.

\begin{table}[]
\centering
\begin{tabular}{ll}\toprule
\multicolumn{2}{l}{\textbf{Indices}}   \\ \hline
$i$           & Container, $i \in \{1, \ldots, n\}$           \\ 
$j$           & Track, $j \in \{1, \ldots, m\}$           \\ \hline

\multicolumn{2}{l}{\textbf{Parameters}} \\ \hline
$c_i^k$     & costs for using multimodal route $k$ for container $i$  \\
$c_i^t$     & costs for using truck by container $i$ \\
$V_j$       & capacity of track $j$ \\
$r_{ik}$     & route $k=1,2,3$ for container $i$ \\ 
$r_{ikj}$     & 
=$ \begin{cases} 
      1 & \text{if route } k \text{ for container } i \text{ contains track } j \\
      0 & \text{otherwise}
   \end{cases}
$ \\ \hline
\multicolumn{2}{l}{\textbf{Decision variables}} \\ \hline
$\{x_{2i-1} \quad x_{2i} \}$     & 
=$ \begin{cases} 
      $\{1 1\}$ & \text{if container } i \text{ is transported by truck} \\
      $\{0 1\}$ & \text{if container } i \text{ is transported by route } r_{i1} \\
      $\{1 0\}$ & \text{if container } i \text{ is transported by route } r_{i2} \\
      $\{0 0\}$ & \text{if container } i \text{ is transported by route } r_{i3} 
   \end{cases}
$ \\ \bottomrule
\end{tabular}
\caption{Parameters and decision variables.}
\label{tab:parameters2}
\end{table}

The binary integer linear program for the problem is formulated as follows:
\begin{align}
\min &\sum_{i=1}^n c_i^3 + (c_i^2-c_i^3)x_{2i-1} + (c_i^1-c_i^3)x_{2i} + (c_i^t+c_i^3-c_i^2-c_i^1)x_{2i-1}x_{2i}, \label{eq:9}\\
\text{such that} \quad &\sum_{i=1}^{n} (1-x_{2i-1})(1-x_{2i}) r_{i3j} + (x_{2i-1}-(x_{2i-1}x_{2i})) r_{i2j} \nonumber \\
&\quad \quad \quad \quad  + (x_{2i}-(x_{2i-1}x_{2i})) r_{i1j} \leq V_j \quad \forall j \in \{1,\dots m\}\label{eq:10}.
\end{align}
\noindent

Here again Eq. \ref{eq:9} minimises the costs of the choices and Eq. \ref{eq:10} makes sure that the capacity constraints of the modalities or tracks are met.

Based on the model in Eq. (\ref{eq:9}-\ref{eq:10}), the QUBO formulation of the problem can also be derived. The overall solution for this QUBO is given by:
\begin{align}
    \min \quad& H=A\cdot H_{A} + B \cdot H_{B} \\
    \text{ with }
    H_{A} &= \sum_{i=1}^n c_i^3 + (c_i^2-c_i^3)x_{2i-1} + (c_i^1-c_i^3)x_{2i} + (c_i^t+c_i^3-c_i^2-c_i^1)x_{2i-1}x_{2i},\\
    H_{B} &= \sum_{j=1}^{m} \Big(\sum_{i=1}^{n} (1-x_{2i-1})(1-x_{2i}) r_{i3j} + (x_{2i-1}-(x_{2i-1}x_{2i})) r_{i2j} \nonumber\\
    &\quad \quad \quad \quad + (x_{2i}-(x_{2i-1}x_{2i})) r_{i1j} + \sum_{k=0}^{K}2^{k}y_{jk} - V_j \Big)^{2},
\end{align}

where $y_{ik}$ denotes additional slack variables. These binary slack variables are necessary to remodel Eq. \ref{eq:10} into equality constraints, as they are required in a QUBO-formulation. Again, $A$, $B$ and $K$ denote penalty coefficients and the number of slack variables per track. They will be set conform the rules explained earlier. 

\subsection{Solving the QUBO problem}
For solving the QUBO problem we use the D-Wave system. Implementing problems on a quantum device asks for specific Quantum Software Engineering \cite{piattini2020talavera}. We will sketch a number of specific implementation issues for this problem.
\subsubsection{Quantum Annealing}
The quantum devices produced by D-Wave Systems are practical implementations of quantum computation by adiabatic evolution \cite{Farhi2000}. The evolution of a quantum state on D-Wave's QPU is described by a time-dependent Hamiltonian, composed of initial Hamiltonian $H_{0}$, whose ground state is easy to create, and final Hamiltonian $H_{1}$, whose ground state encodes the solution of the problem at hand:
\begin{align}\label{eq:Hamiltonian}
   H(t) = \Big(1-\frac{t}{T}\Big) H_{0}+\frac{t}{T} H_{1}.
\end{align}
The system in Eq.~\eqref{eq:Hamiltonian} is initialised in the ground state of the initial Hamiltonian, i.e., $H(0)=H_{0}$. The adiabatic theorem states that if the system evolves according to the  Schr\"odinger equation, and the minimum spectral gap of $H(t)$ is not zero, then for time $T$ large enough, $H(T)$ will converge to the ground state of $H_{1}$, which encodes the solution of the problem. This process is known as quantum annealing. Goal is to find the eigenstates of $H$, where the eigenvalues are the energy values. The eigenstates with the lowest eigenvalue are the ground states of the system. Although here we are not concerned with the technical details, it is worthwhile to mention that it is not possible to estimate an annealing time $T$ to ensure that the system always evolves to the desired state. Since there is no estimation of the annealing time, there is also no optimality guarantee.

The D-Wave quantum annealer can accept a problem formulated as an Ising Hamiltonian, corresponding to the term $H_{1}$ in Eq.~\eqref{eq:Hamiltonian}, or rewritten as its binary equivalent, in QUBO formulation. Next, this formulation needs to be embedded on the hardware. In the most developed D-Wave 2000Q version of the system, the 2048 qubits are placed in a Chimera architecture\cite{mcgeoch2014adiabatic}.: a $16\times 16$ matrix of unit cells consisting of $8$ qubits. This allows every qubit to be connected to at most 5 or 6 other qubits. With this limited hardware structure and connectivity, fully embedding a problem on the QPU can sometimes be difficult or simply not possible. In such cases, the D-Wave system employs built-in routines to decompose the problem into smaller sub-problems that are sent to the QPU, and in the end reconstructs the complete solution vector from all sub-sample solutions. The first decomposition algorithm introduced by D-Wave was \emph{qbsolv} \cite{qbsolv}, which gave a first possibility to solve larger scale problems on the QPU. Although \emph{qbsolv} was the main decomposition approach on the D-Wave system, it did not enable customisations, and therefore is not particularly suited for all kinds of problems. The new decomposition approaches D-Wave offers are D-Wave Hybrid and the Hybrid Solver Service, offering more customisations.\\

Next to the QPU also a CPU can be used within the programming environment of D-Wave, using for example simulated annealing (SA), a conventional meta-heuristic.  

\subsection{Embedding}
Some restrictions are introduced by the hardware design of the D-Wave hardware. The current implementation has a qubit connectivity based on a Chimera structure. The QUBO problem at hand has to be transformed to this structure. Because of the limited chip sizes we have currently, a compact formulation of the QUBO is important, but also a compact transformation to the chip design. This problem is known as minor-embedding. Here the vertices correspondent to problem variables and edges exist if $Q_{ij} \neq 0$, where $Q$ represents the qubo-matrix. Because of the limited connectivity of the chip, a problem variable has to be duplicated to multiple (connected) qubits. Those qubits should have the same value, meaning the weight of their connection should be such that it holds in the optimisation process. All these qubits representing the same variables are part of a so-called chain, and their edge weights is called the chain strength ($\lambda$), which is an important value in the optimisation process. 
In \cite{coffrin}, Coffrin gives a thorough analysis. He indicates that if $\lambda$ is sufficiently large, optimal solutions will match $\lambda \geq \sum_{ij}|Q_{ij}|.$ However, the goal is to find the smallest possible value of $\lambda$, to avoid re-scaling of the problem. Coffrin also indicates that finding the smallest possible setting of $\lambda$ can be NP-hard. Also other research has been performed on selecting an optimal choice for this chain strength \cite{rieffel2015case}, but at the moment there is no solid criterion for choosing a value. A rule of thumb that is suggested\footnote{\url{https://support.dwavesys.com/hc/en-us/community/posts/360034852633-High-Chain-Break-Fractions}} is $\lambda = \max_{ij} Q_{ij}$. It may be necessary to use the quantum annealer with multiple values of the chain strength in order to determine which value for $\lambda$ is optimal for a given problem \cite{foster2019applications}.  

\section{Numerical Results}
\subsection{2-Alternative Route Case Study}
We use a case consisting of 10 containers ($n=10$) and 12 tracks ($m=12$). The parameters are given in Table \ref{tab:para}.
\begin{table}[]
    \centering
    \begin{tabular}{c c}
    \toprule
\multicolumn{2}{c}{$c_b = (2 ,7 ,1 ,6 ,2 ,4 ,8 ,7 ,7 ,10 )$}\\ 
\multicolumn{2}{c}{$c_t = (23 ,25 ,23 ,17 ,24 ,22 ,19 ,16 ,21 ,17 )$}\\
\multicolumn{2}{c}{$v = (5 ,5 ,5 ,5 ,5 ,5 ,5 ,5 ,5 ,5 ,5 ,5 )$} \\
&\\
$r_1 = (1,0,1,0,0,1,0,1,0,0,0,0) $ &
$r_2 = (1,0,1,0,1,0,1,0,0,0,0,0) $ \\
$r_3 = (1,0,1,0,0,1,0,1,0,0,0,0) $ &
$r_4 = (1,0,0,0,0,0,1,0,1,1,0,0) $ \\
$r_5 = (1,0,1,0,1,0,1,0,0,0,0,0) $ &
$r_6 = (0,1,0,1,0,1,0,1,0,0,0,0) $ \\
$r_7 = (1,0,1,0,1,0,1,0,0,0,0,0) $ &
$r_8 = (1,0,1,0,1,0,1,0,0,0,0,0) $ \\
$r_9 = (1,0,1,0,1,0,1,0,0,0,0,0) $ &
$r_{10} = (0,1,0,1,0,1,0,1,0,0,0,0) $\\
&\\
\toprule
    \end{tabular}
    \caption{Parameters used in the 2-alternative route case study.}
    \label{tab:para}
\end{table}

In the optimal solution, containers 4, 7 and 8 are transported by trucks and the others by barges, resulting in a total costs of 85, using the standard Matlab solver. As explained, this size of problems is no problem for classical solvers. The goal of this exercise is to show th potential of quantum computing when the quantum computers have more mature sizes. The QUBO following from Eq. \ref{eq:6} leads to a 46x46 matrix.

\subsection{Simulated Annealing}
\begin{figure}
    \centering
    \includegraphics[width=12cm]{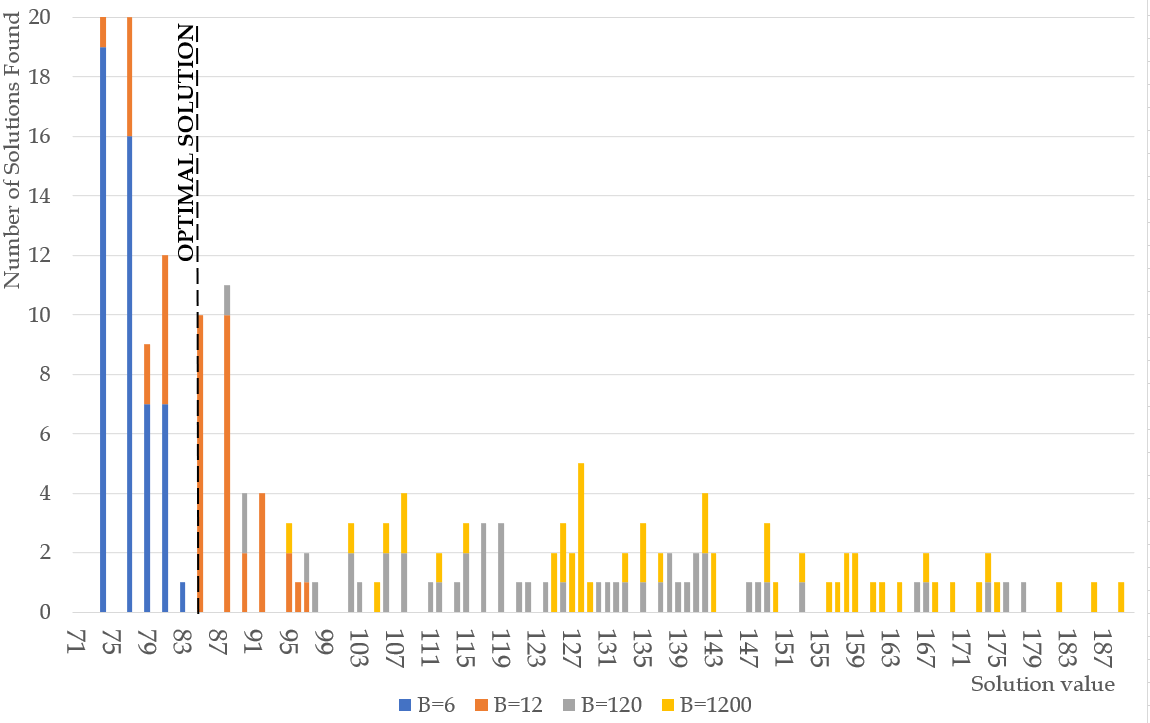}
    \caption{Results for four values of $B$ for the simulated annealing approach.}
    \label{fig:sim1}
\end{figure}

The annealing processes (both simulated as quantum) are stochastic processes. This means that if we do a high number of queries we get a probability distribution for the resulting solution. The first important parameter to fix is this number of queries. We chose 500 samples. Next the values for $A$ and $B$ are important. We can set $A=1$ and look for a good value for $B$. Rule of thumb is that the gain of violating a constraint must be higher than the consequent reduction in cost. This means in this problem that $B>c_b$, so $B>10$. In Figure \ref{fig:sim1} the probability distribution of four values for $B$. We see that choosing $B$ too low ($B=6$) gives all solutions that are better than the optimal solution, but they all violate one or more constraints. Choosing $B$ too high gives solutions that are feasible, but they are (far) away of the optimal solution. Choosing $B=12$ gives a range of solutions that include the optimal value, which will be found in most runs. The computation time for approach is around one minute, much more than conventional solvers will ask, who aks for less than one second.

\subsection{Quantum Annealing}
To use D-Wave's QPU now two parameters have to be determined. Again the value for $A$ and $B$ are important. By setting $A=1$ we can choose a good value for $B$. Here we expect that the same value as established for simulated annealing will work. Here, we also have to determine a suitable value for the chain strength $\lambda$. We start with the rule of thumb indicated in the previous section, $\lambda = \max_{ij} Q_{ij}$. Note that Eq. \ref{eq:1}-\ref{eq:3} imply that $\lambda$ is dependent on the value of $B$, because $B$ is part of the $Q$ matrix.  If we choose the values $3$, $6$ and $12$ for $B$, then the values $120$, $240$ and $480$ follow for $\lambda$. If we use $\lambda \geq \sum_{ij}|Q_{ij}|,$ the values $4,673$, $9,341$ and $18,687$ follow. In Table \ref{tab:results} the results from the annealing are depicted. For each combination $\{\lambda,B\}$ we present the minimum value, the average value and the percentage of feasible solutions found using $1000$ readouts. The minimum value comes from the solution with the lowest energy that does not violate any of the constraints. The average value is the average objective value of all solutions that do not violate any of the constraints.  

\begin{table}[]
    \centering
    \begin{tabular}{cccccccccccc}
    \toprule
& \multicolumn{10}{c}{B}\\
\cline{2-12}
Chain &	\multicolumn{3}{c}{3}&&\multicolumn{3}{c}{6}	&&\multicolumn{3}{c}{12}\\	
Strenght & min. & avg. & perc. && min. & avg. & perc. && min. & avg. & perc. \\
\cline{2-4} \cline{6-8} \cline{10-12}
1	&85&111&52\%	&&92&135&96\%	&&128&167&100\%	\\
2	&85&103&5\%	&&96&129&50\%	&&134&139&97\%	\\
5	&99&110&2\%	&&98&120&11\%	&&104&124&72\%	\\
10	&98&98&1\%	&&100&117&18\%	&&129&120&40\%	\\
120	&92&118&58\%	&&104&128&87\%	&&92&126&93\%\\	
240	&90&145&93\%	&&115&130&83\%	&&106&124&80\%\\	
480	&96&147&95\%	&&131&135&93\%	&&95&146&95\%\\
4673 &110&149&85\%\\
9341 &&&&& 142 & 153& 98\%\\
18,687 &&&&&&&&& 158 & 160 & 99\%\\
    \toprule
    \end{tabular}
    \caption{Results for Quantum Annealing for various combinations of the chain strength and the penalty value $B$. Each result shows the minimum value, average value, and percentage of allowed solutions.}
    \label{tab:results}
\end{table}

\begin{figure}
    \centering
    \includegraphics[width=12cm]{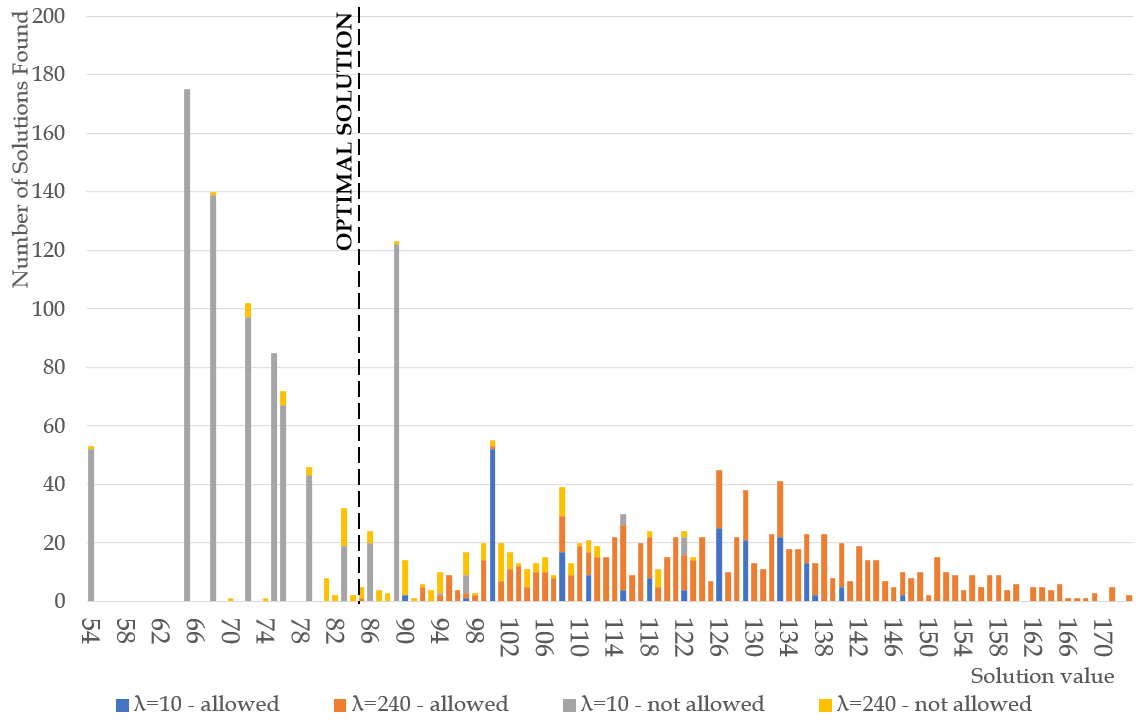}
    \caption{Results for $B=6$ and two values for $\lambda$ for the quantum annealing approach.}
    \label{fig:sim2}
\end{figure}

Note in Table \ref{tab:results} that the best (the optimal) solution is found in the left upper corner, for low values for both $\lambda$ and $B$. The suggested value by the rule of thumb does not give good solutions in this example nor do the upper bound values. We also see that the best average values appear at a value $\lambda=10$ for all values for $B$. This value for $\lambda$ also comes with the lowest percentage feasible solutions, indicating that the probability distribution, like we see in Figure \ref{fig:sim1}, shifts to the left for these values of $\lambda$. For two values ($\lambda=10$ and $\lambda=240$ and $B=6$ the distribution is plotted in Figure \ref{fig:sim2}, confirming this idea. Note that for $\lambda=240$ the best solution found, based on minimum energy, has objective value 115. In Figure \ref{fig:sim2} we see that much better feasible solutions were found (even the optimal solution). These solutions follow from solution that have a chain break(s) and have a higher energy value. Same holds for $\lambda=10$. Here the solution with objective value $90$ is found, where $100$ is reported as best solution. Calculation time of this approach is around 10 ms (pure QPU time) for each readout, leading to 10 second for 1000 readouts. This is more than conventional solvers ask for this size of problems, however, the idea is that Quantum Annealing will scale better for bigger problems, when the hardware is ready for that size of problems. 

\section{Conclusions}
The rise of quantum computing and the promising application to combinatorial optimisation ask for re-formulation of many real-world problems to match the form these quantum computer can handle, the QUBO formulation, where QUBO stands for Quadratic Unconstrained Binary Optimisation. In this paper we proposed two QUBO-formulations for multimodal container planning, where sets containing a limited number of paths are used. For each container an assignment is found that minimises the overall costs of transporting all containers to their destination. These formulations are proposed to sketch to potential of quantum computing for this kind of problems when the quantum computer will be a more matured technology.

When implementing these problem formulations on D-Wave's quantum annealer, there are a number of implementation issues such as finding the right embedding, defining the chain strength and finding the right penalty functions. We showed how this is addressed by doing a parameter grid search.

For further research a more general idea for finding these parameters is recommended. Also the extension of the set with alternative paths is recommended and a integrated way to include promising paths to this set. Here we think of introducing a quantum variant of the known classical technique of column generation, Quantum Column Generation.

\bibliographystyle{splncs04}
\bibliography{main.bib}

\section*{Appendix: Code example}
\begin{Verbatim} [fontsize=\small,breaklines=true]
import neal
import numpy 
from pyqubo import Array, Placeholder, solve_qubo, Constraint
from pyqubo import Sum, Model, Mul
from dwave.system.samplers import DWaveSampler           
from dwave.system.composites import EmbeddingComposite   

# Initialize parameters 
N = 10 # number of containers 
M = 12 # number tracks 
K = 3  # slack parameter
 
# Costs 
c_b = [2,7,1,6,2,4,8,7,7,10] 
c_t = [23,25,23,17,24,22,19,16,21,17] 
 
# Track capacity 
v = {} 
v = [5,5,5,5,5,5,5,5,5,5,5,5] 
 
routes = {} # route of containers 
routes[0] = [1,0,1,0,0,1,0,1,0,0,0,0] 
routes[1] = [1,0,1,0,1,0,1,0,0,0,0,0] 
routes[2] = [1,0,1,0,0,1,0,1,0,0,0,0] 
routes[3] = [1,0,0,0,0,0,1,0,1,1,0,0] 
routes[4] = [1,0,1,0,1,0,1,0,0,0,0,0]
routes[5] = [0,1,0,1,0,1,0,1,0,0,0,0] 
routes[6] = [1,0,1,0,1,0,1,0,0,0,0,0] 
routes[7] = [1,0,1,0,1,0,1,0,0,0,0,0] 
routes[8] = [1,0,1,0,1,0,1,0,0,0,0,0] 
routes[9] = [0,1,0,1,0,1,0,1,0,0,0,0] 

# Initialize variable vector
size_of_variable_array = N + K*M
var = Array.create('vector', size_of_variable_array, 'BINARY')

# Defining constraints in the Model
minimize_costs = 0
minimize_costs += Constraint(Sum(0, N, lambda i: var[i]*(c_t[i]-c_b[i])+c_b[i]),label="minimize_transport_costs")
capacity_constraint = 0
for j in range(M):
    capacity_constraint += Constraint ( (Sum(0, N, lambda i: (1-var[i])*routes[i][j]) + Sum(0, K, lambda i: var[N + j*K + i]*(2**(i))) - v[j])**2, label= "capacity_constraints")
    
# parameter values
A = 1
B = 6
Cs = 240
useQPU=True
    
# Define Hamiltonian as a weighted sum of individual constraints
H = A * minimize_costs +  B * capacity_constraint

# Compile the model and generate QUBO
model = H.compile()
qubo, offset = model.to_qubo()

# Choose sampler and solve qubo
if useQPU: 
    sampler = EmbeddingComposite(DWaveSampler())
    response = sampler.sample_qubo(qubo, chain_strength=Cs, num_reads = 1000) #solver=DWaveSampler()) #, num_reads=50)   
else:
    sampler = neal.SimulatedAnnealingSampler()
    response = sampler.sample_qubo(qubo, num_sweeps=10000, num_reads=50)   

# Postprocess solution
sample  = response.first.sample
\end{Verbatim}

\end{document}